# Characterizing TMS-EEG perturbation indexes using signal energy: initial study on Alzheimer's Disease classification

Alexandra-Maria Tăuțan[1,2,*], Elias Casula[3], Ilaria Borghi[3], Michele Maiella[3], Sonia Bonni[3], Marilena Minei[3], Martina Assogna[3], Bogdan Ionescu[1], Giacomo Koch,[3,4] and Emiliano Santarnecchi[2]

*Abstract*-Transcranial Magnetic Stimulation (TMS) combined with EEG recordings (TMS-EEG) has shown great potential in the study of the brain and in particular of Alzheimer's Disease (AD). In this study, we propose an automatic method of determining the duration of TMS- induced perturbation of the EEG signal as a potential metric reflecting the brain's functional alterations. A preliminary study is conducted in patients with Alzheimer's disease (AD). Three metrics for characterizing the strength and duration of TMS-evoked EEG (TEP) activity are proposed and their potential in identifying AD patients from healthy controls was investigated. A dataset of TMS-EEG recordings from 17 AD and 17 healthy controls (HC) was used in our analysis. A Random Forest classification algorithm was trained on the extracted TEP metrics, and its performance is evaluated in a leave-one-subject-out cross-validation. The created model showed promising results in identifying AD patients from HC with an accuracy, sensitivity, and specificity of 69.32%, 72.23% and 66.41%, respectively.

*Clinical relevance*- Three preliminary metrics were proposed to quantify the strength and duration of the response to TMS on EEG data. The proposed metrics were successfully used to identify Alzheimer's disease patients from healthy controls. These results proved the potential of this approach which will provide additional diagnostic value.

## I. INTRODUCTION

Transcranial Magnetic Stimulation (TMS) co-registered with electroencephalography (EEG) shows great potential in the study of the healthy and pathological brain. By inducing a strong and focused magnetic field, electrical currents are induced into targeted regions of the brain [1]. This in turn creates local and global TMS-evoked EEG potentials (TEPs) that allow to investigate the reaction of specific brain regions to external perturbation [2].

There is currently no clear consensus on the methods to be used for TEP analysis. Several latencies and polarities have been analyzed with respect to diverse pathologies [3], [4], but no consensus exists on when and how long a response to TMS is expected. The strength and duration of the response to stimulation can be indicative of specific pathologic conditions where neuronal pathways and responses are altered.

For instance, Alzheimer 's disease (AD) is a neurodegenerative disorder in which aberrant proteins accumulate both intra and extracellularly in different regions of the brain [5].This causes altered neuronal behaviors which induce changes in brain networks ' dynamics, excitability and interneuron communication [6]. Its prevalence worldwide is high, and it is expected to double by 2060, while currently no effective treatments are available [7]. TMS-EEG can be a great tool in the study of this disease as both local and distant, network level responses can be analyzed.

Most works using TMS-EEG to study AD focus on the early TEP responses, within l00ms after the stimulation. Little information is available on the later TEP responses. A common assumption is that the immediate effect of TMS perturbation stops roughly after 500ms or even earlier [8]. However, TEPs also show high inter-subject variability. We assume that the strength and duration of the response is different for each subject. Furthermore, we hypothesize that the response to stimulation in both strength and duration is different between AD patients and healthy controls.

In this study we aim to: (i) investigate an energy-based method of determining the return to baseline after TMS stimulation, (ii) propose metrics for quantifying the response to stimulation, and (iii) investigate the possibility of classifying AD from HC using the proposed metrics.

This paper is organized as follows. Section II provides an overview of the classification algorithm for identifying AD from HC. Section III details the method used for automatically determining the time the EEG returns to baseline after perturbation, along with other metrics to quantify the perturbation strength. Section IV presents and discusses the results, and Section V concludes the paper.

## II. METHODS

An overview of the general framework and steps for classifying AD patients from HC is provided in Figure 1. The TMS-EEG dataset is pre-processed prior to further analysis. Features based on the return to baseline time are extracted for each individual trial and averaged over the entire electrode set per patient. The obtained values are used as input for classification. The created model is in the end evaluated to quantify performance.

[1] Alexandra-Maria Tăuțan and Bogdan Ionescu is with University Politehnica of Bucharest, AI Multimedia Lab, Research Center CAMPUS, 6 luliu Maniu Bd., 061344, Bucharest, Romania (e-mail: bogdan.ionescu@upb.ro)

[2] Alexandra-Maria Tăuțan and Emiliano Santarnecchi are with Preci- sion Neuromodulation Program Network Control Laboratory, Gordon Center for Medical imaging, Department of Radiology, Massachusetts General Hospital, Harvard Medical School, Boston, MA, USA (e- mail:atautan@mgh.harvard.edu, esantarnecchi@mgh.harvard.edu)

[3] Ilaria Borghi, Michele Maiella, Sonia Bonni, Marilena Minei, Martina Assogna, Elias Casula and Giacomo Koch are with the Santa Lucia Foundation, Via Ardeatina 306354, 00179, Rome, Italy (e- mail: e.casula@hsantalucia.it, g.koch@hsantalucia.it)

[4] Giacomo Koch is with Department of Neuroscience and Rehabilitation, Section of Human Physiology, University of Ferrara, Via Fossato di Mortara 17-19, 44121 Ferrara, Italy

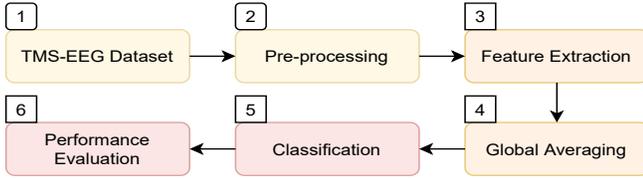

Fig. 1. Steps in creating a classification model for identifying AD patients from HC.

## A. Dataset

EEG activity from the scalp was collected during a TMS protocol with a BrainAmp amplifier (BrainProducts GmbH, Munich, Germany). A montage of 64 TMS-compatible Ag/AgCl electrodes was used. A figure of eight coil oriented at a 45° angle from the midline was used with a Magstim Rapid magnetic biphasic stimulator to apply TMS to the left dorsolateral prefrontal cortex. The stimulation intensity was determined based on a distance-adjusted motor threshold dependent on the individual coil-to-cortex distance [9]. Blocks of 120 single pulses were applied with an inter-stimulus interval of 1 to 4 seconds.

TMS-EEG data from 17 AD patients and 17 healthy aged-matched controls are used for the experiments. The average age of the two groups is 72.35±7.72 and 71.11±6.28 respectively. All participants were assessed for signs of cognitive decline using clinical and neuropsychological data based on the latest AD diagnostic criteria [10].

The dataset was collected at the Santa Lucia Foundation (Rome, Italy). The study was approved by the ethics committee of the Santa Lucia Foundation and was conducted according to the principles of the Declaration of Helsinki and the International Conference on Harmonization of Good Clinical Practice. Written consent was obtained from all participants, or their legal representatives and they were informed participation is voluntary.

## B. EEG Preprocessing

EEG data collected while apply TMS is subject to multiple sources of artefacts. Preprocessing was performed on the data to eliminate confounding artefacts (Fig. 1, Step 2) [11]. The EEG was segmented 500ms prior and 1000ms after the TMS pulse. The recording during the TMS pulse was removed and a cubic interpolation was performed. A zero-phase Butterworth bandpass filter was applied between 1 and 80Hz on the data down sampled to 1000Hz. Several independent component analyses (ICA) were performed with manual selection and removal of remaining interfering factors [12]. The electrode channels were re-referenced to the electrode average.

## C. Classification and Evaluation

After preprocessing, features were extracted on individual TMS trials (Fig. 1, Step 3). The features extracted were exclusively related to the timepoint determined for the EEG signal returned to the values from the baseline. These are further detailed in Section III. The values obtained are averaged over all trials and over all electrodes for each subject (Fig. 1, Step 4). The obtained feature set is used as input for creating a classification model to distinguish between AD and HC (Step 5). A Random Forest algorithm with 100 trees and a minimum of 1 sample per leaf was used for creating the mode. Random Forest is an ensemble learning technique that uses majority voting from multiple decision trees to create a final decision on classification. Each decision tree receives as input a subpart of the dataset and thus eliminating the problem of overfitting in the final model [13].

A leave-one-out cross-validation method was used (Fig. 1, Step 6). In this method, $k-1$ subjects are used for training the model and the $k^{th}$ subject is used as a test. To account for variations in bootstrapping the original data, the algorithm is run 100 times and the average result over all runs is reported. The metrics defined in equations 1-4 below are used for evaluating the performance of the model:

$$\text{accuracy} = (TP+TN)/(FP+FN+TP+TN) \quad (1)$$

$$\text{sensitivity} = TP/(TP+FN) \quad (2)$$

$$\text{specificity} = TN/(TN+FP) \quad (3)$$

$$F_1 \text{ score} = 2TP/(2TP+FP+FN) \quad (4)$$

where $TP$ - true positive, $TN$ - true negative, $FP$ - false positive, $FN$ - false negative.

## III. DETERMINING THE RETURN TO BASELINE

### A. Signal Energy

Unlike other event related potentials such as visual or audio ERPs that have an expected polarity, amplitude increase or decrease at specific latencies after stimulus, the TEP shows a high inter-subject variability in the response to TMS stimulation both in time and frequency [14]. Several proposals for latency-based analysis have been made in literature [3], [4]. However, no consensus exists on the type of analysis that should be applied to TMS-induced potentials. To overcome this shortcoming, metrics such as area under the curve were used to quantify the strength of the response to TMS stimulation [15].

Here, we use signal energy to characterize the EEG response to TMS stimulation. The median energy of a signal is defined according to equation 5:

$$Energy = \frac{\sum_i^N |x_i|^2}{N} \quad (5)$$

where $x_i$ is a sample and $N$ is the total number of samples in the evaluated signal segment.

Immediately after the TMS pulse is applied, the EEG signal energy increases while exhibiting a slower or faster decrease over time as the immediate effects of perturbation fade out [8]. Our assumption is that the signal energy after TMS decreases until the levels observed during baseline are obtained.

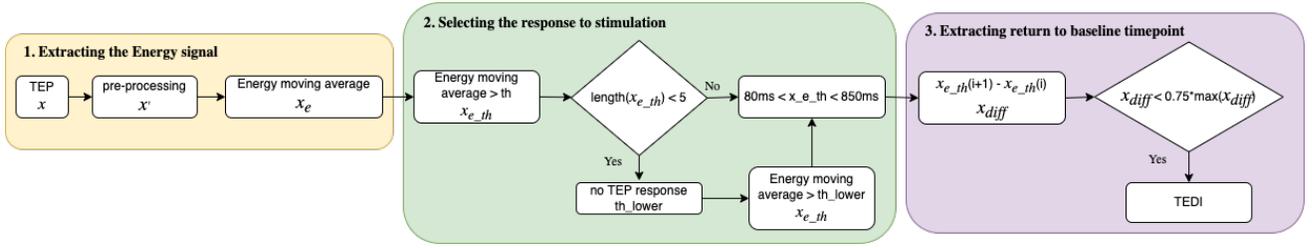

Fig. 2. Block diagram of the algorithm automatically determining the return to baseline of the EEG signal after applying TMS.

*B. Algorithm proposal*

Our proposal for automatically calculating the time needed for the EEG signal to settle back to baseline is based on the signal energy levels obtained over the time course of the TEP. Figure 2 shows the block diagram of our algorithm proposal. Three main blocks are identified as follows:

*1) Extracting the energy signal:* The absolute TEP value is extracted and normalized with respect to its maximum value to have a uniformity in amplitude ranges in-between subjects. Next, the energy level of the signal is calculated over a sliding window moving sample by sample through the TEP. Experiments were performed to evaluate the optimum window size. Values for the length of the sliding window included 5ms, l0ms, 20ms, 30ms and 40ms. The final evaluation was conducted with respect to the maximum performance obtained for the classification of AD and HC.

*2) Selecting the response to stimulation:* Once the energy signal $X_e$ is calculated, the values obtained after the TMS pulse are compared to baseline. A threshold is defined as the mean value of the energy signal obtained on the baseline data added with one standard deviation. Baseline has been considered between -500ms and -200ms prior to the pulse. The last 200ms prior to the pulse were excluded to avoid the interference of residual pre-processing and filtering effects on the baseline data. After applying the threshold, a new signal $X_{e\_th}$ is obtained that should be representative of the increase in energy as a response to stimulation. The time points where the threshold was crossed by $X_e$ are also retained. Several control measures are put in place. If less than 5 samples are available in $X_{e\_th}$, the threshold is lowered to the mean of the baseline to ensure a return to baseline value can be calculated. The time interval where the return to baseline can occur is limited between 80ms and 850ms. The lower bound assumes that there always is a response to stimulation until 80ms. The upper bound takes into account previous literature where generally the return to baseline is considered at approximately 500ms after stimulation [8].

*3) Extracting return to baseline time point:* The difference between the timepoints when the energy signal crossed the threshold $X_{diff}$ is computed. The maxim interval between time points is extracted. When the difference between two time points of the $X_{e\_th}$ signal is higher than 75% of the maximum of $X_{diff}$, it is considered that the signal is settled after the perturbation. In our definition, this represents the time required for the signal to return to baseline.

*C. Metrics*

Three metrics to characterize the TEP are proposed. These are detailed and defined in Table I. Figure 3 also provides a graphical view of the metrics and how they are determined from an example TEP. The time required for the EEG signal to return to baseline after applying TMS can be useful to characterize the ability of the brain to recover after perturbation, possibly indexing plasticity mechanisms as well. However, the return time is not the only indicator of recovery. For instance, AD patients have been previously shown to exhibit increased excitability [6]. Therefore, the combination between the amplitude of the response and the time required to return to baseline energy could reveal important information on the pathology and could serve as a biomarker. We propose two metrics to measure this relationship: a measure of the decay in energy and the area under the curve of the detected response.

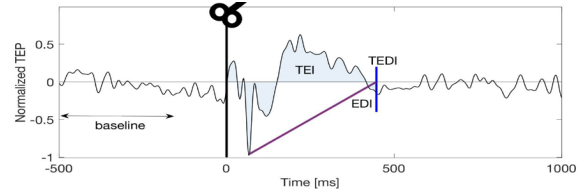

Fig.3. Overview of features extracted from the TEP based on the time required for the EEG to return to baseline.

TABLE I
OVERVIEW OF METRICS USED IN CLASSIFICATION ALONG WITH THEIR ABBREVIATION (ABB.) AND THEIR DEFINITION.

| Metric | Abb. | Definition |
| --- | --- | --- |
| Target engagement duration index | TEDI | Time required for the signal to return to baseline. See Section lll-B |
| Engagement decay index | EDI | Angle between TEDI and the maximum absolute value of the TEP between the TMS pulse and TEDI |
| Target engagement index | TEI | Area under the curve of the signal between the TMS pulse and TEDI |

## IV. RESULTS AND DISCUSSION

The initial experiments showed a good classification performance between AD and HC was obtained using the metrics defined based on the time required for the EEG to revert back to baseline levels after TMS stimulation. Figure 4 shows the classification results in terms of accuracy, sensitivity, specificity, and $F_1$ score when varying the sliding window size to determine TEDI. The most stable results along all metrics are obtained using a sliding window of 20ms. The accuracy, sensitivity, specificity and $F_1$ score were of 69.32%, 72.23%, 66.41% and 69.27% respectively.

Our results indicate that the proposed metrics show potential in differentiating between Alzheimer's disease patients and healthy controls. However, the performance is lower when compared to AD classification models obtained from, for instance resting state EEG [16]. This is most likely due to the preliminary nature of the study and the fact that only three features characterizing TEPs have been used as input. Additional time and frequency domain features could increase the classification performance. Furthermore, frequency content alterations are well known in resting state EEG recordings [17]. Perturbation metrics that also encode frequency content could be as well useful in determining and characterizing the response to TMS from the EEG signal.

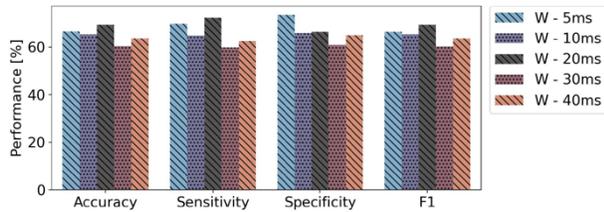

Fig. 4. Performance classification based on the variation of the length of the sliding window used to determine the return to baseline.

TMS-EEG shows significant inter-subject variability. While some extreme responses might be observed because of pathology, some patients might show reduced or no responses to stimulation. The proposed algorithm does not detect trials that might in fact exhibit no response to stimulation as the threshold is lowered once the energy signal does not cross the threshold sufficiently (Figure 2, block 2). Further investigation is needed to determine the optimum threshold choice and if these metrics can be used to also evaluate the lack of responses to TMS.

Finally, the proposed metrics to characterize the response to TMS were evaluated based on their performance in the problem of AD identification. However, the response to TMS of AD patients is still a topic of active research. Further validation would be needed with respect to established measures of quantifying subject response to stimulation such as motor evoked potentials [18] or increase in heart rate responses. These measures should also be tested longitudinally as potential metrics for disease tracking, in conjunction with behavioral and cognitive data.

## V. CONCLUSION

In this work we have proposed three metrics to quantify the response to TMS stimulation on EEG data based on signal energy. Results showed a good classification performance of 69.32%, 72.23% and 66.41% accuracy, sensitivity, and specificity respectively. Regardless of the magnitude of classification performance, analysis of signal energy represents an innovative approach to quantify and model the brain's response to perturbation, with strong potential for characterizing disease states.

Future work will investigate other signal characteristics that could indicate the presence of a response to TMS stimulation and will validate the metrics with respect to otherperturbation measures extracted from the motor system.